\documentclass[12pt]{article}
\usepackage{epsfig}

\usepackage{pdfpages}
\def\beq{\begin{equation}}
\def\eeq#1{\label{#1}\end{equation}}
\def\eeqn{\end{equation}}

\def\beqa{\begin{eqnarray}}
\def\eeqa#1{\label{#1}\end{eqnarray}}
\def\eeqan{\end{eqnarray}}

\let\bar=\overbar

\def\Dslash{\not{\hbox{\kern-4pt $D$}}}
\def\dslash{\not{\hbox{\kern-2pt $\del$}}}

\def\msb{{\bar{\ssstyle M \kern -1pt S}}}

\def\Title#1{\begin{center} {\Large {\bf #1} } \end{center}}

\begin{document}

\Title{Future Prospects at e$^+$e$^-$ Machines}

\bigskip\bigskip


\begin{raggedright}  

{\it J. Michael Roney\index{Roney, J.M.}\\
Department of Physics and Astronomy\\
University of Victoria\\
Victoria, BC, CANADA}
\bigskip\bigskip
\end{raggedright}

\section{Introduction}

This review will present the physics prospects for physics at e$^+$e$^-$
machines running  near the $\Upsilon(4S)$ as well as at the charm and tau thresholds.
I will briefly review the plans for BES~III, which  operates near the charm threshold
at  IHEP's BEPC machine in Bejing and the status of the  Super Charm/Tau Factory proposed
for the Budker Institute for Nuclear Physics (BINP) in Novosibirsk.

At the time of this conference the particle physics community was looking forward to
having two next generation high luminosity e$^+$e$^-$ machines running at or near
the $\Upsilon(4S)$ operating by the end of the decade:
the Belle~II experiment at the SuperKEKB machine in Japan and the SuperB experiment
at the recently created Cabibbo Laboratory in Italy.
 Before the end of 2012, however, the unfortunate
fiscal difficulties facing many European countries forced the Italian government to cancel SuperB.
Nonetheless, I will discuss both of these projects in this review.

\section{BES~III Datasets and Future Plans}
In 2009 BES~III collected 106 million events on the $\psi(2S)$ resonance and 225 million 
at the J/$\psi$ resonance, where each represents a sample that is four times larger 
 than the samples collected by CLEO-c
and BES~II, respectively. The program in 2010 and 2011 focused on running at the $\psi(3770)$
and $\psi(4040)$ where a total of 2700~pb$^{-1}$ and 470~pb$^{-1}$ of data were collected, respectively.
The physics running in 2012 saw the accumulation of data at tau-pair threshold to be used in a new
precision tau mass measurement, 400 million events at the $\psi(2S)$, one billion J/$\psi$ events,
and an energy scan to measure R, the ratio of the total hadronic to muon-pair cross sections.

Looking to the future, BES~III will collect data at centre-of-mass energies of 4260~MeV and
4360~MeV in order to conduct spectroscopic studies of the  ``XYZ''  states and they will also
continue the R scan. Plans for 2014 have the machine running at 4170~MeV in order to collect
roughly 2.4~fb$^{-1}$ of data to study the D$_s$. There are no firm plans for the 
program beyond 2014, although the collaboration would like to collect 10~fb$^{-1}$ of $\psi(3770)$
data. BES~III is scheduled to collect data for another eight to ten years.

\section{BINP Super Charm/Tau Factory}
The physics goals of the Super Charm/Tau Factory being planned
for Novosibirsk~\cite{Levichev:2008zz}  includes:
\begin{itemize}\item High statistics spectroscopy and searches for exotics
 \begin{itemize}
   \item Charm spectroscopy
   \item Spectroscopy of the highly excited charmonium states (complementary to bottomonium)
   \item Light hadron spectroscopy in charmonium decays
 \end{itemize}
\item Precision charm physics
 \begin{itemize}
   \item Precision charm CKM (strong phases, $f_D$, $f_{D_s}$,form factors, etc)
   \item Use this unique source of coherent $D^0$/$\overline{D^0}$ states for a variety of measurements
       ($D^0$ mixing, CPV in
         mixing, strong phases for $\gamma$/$\psi_3$ measurements at SuperB/Belle~II and LHC)
 \end{itemize}
\item Precision $\tau$-physics with polarized beams
\begin{itemize}
 \item Lepton universality, Lorentz structure of $\tau$-decay
 \item CP and T-violation in $\tau$ and $\Lambda_C$ decays
 \item Search for lepton-flavour-violating decays such as $\tau\to\mu\gamma$
 \item Measure second class currents (with kinematical constraints at threshold)
\end{itemize}
\item Two photon physics and light hadronic cross section via ISR
\end{itemize}

The machine will operate at
centre-of-mass energies  ranging from 1.0 to 2.5~GeV with online energy monitoring
with a precision at the level of 5-10$\times 10^{-5}$ and
a design peak luminosity of 10$^{35}$~cm$^{-2}$s$^{-1}$ at 2~GeV.
Also, the electrons are to be polarized longitudinally at the interaction point.

The collider will have two separate rings utilizing the crab waist
 collision scheme\cite{Levichev:2008zz} at a single interaction point (IP) with 
a goal of a reaching sub-millimeter beta-y at the IP. A 2.5~GeV linac will
 be used to inject beam at full energy. 
Four super conducting wigglers will be used to help
preserve damping parameters through the whole energy range in order
 to optimize the luminosity. Five Siberian snakes will be
used to obtain the longitudinally polarized electrons, again
 for the whole energy range.
It will use a polarized electron source and a positron source
that will have a 50~Hz top-up injection frequency.

Although the project was included in the 2012 list of the
top six projects approved for further development by
the Russian Governmental Commission on the
Innovations and High Technologies, at this time the project is
not yet approved for funding.

\section{The Super B Factory Experiments: Belle~II and SuperB}

These facilities are designed to collide electrons and positrons at  centre-of-mass
energies near the $\Upsilon$ resonances. Most of the data will be collected
at the $\Upsilon(4S)$ resonance, which is just above threshold for B-meson pair production
where no fragmentation particles are produced. The luminosity 
is expected to be 40-100 times that of  the previous generation colliders:
SuperKEKB~\cite{Abe:2010sj} has a design luminosity of $8\times 10^{35}$cm$^{-2}$s$^{-1}$
whereas SuperB~\cite{Bona:2007qt} was designed for  $10^{36}$cm$^{-2}$s$^{-1}$.
These would produce  5-10$\times 10^{10}$ b, c and tau pairs 
corresponding to 50-75~ab$^{-1}$. Both machine designs
have asymmetric beam energies in order to give a boost to
the centre-of-mass system and thereby allow for time-dependent CP
violation measurements.

\subsection{Physics Program}
The overall physics goal of these next generation flavour factories 
is to search for evidence of new physics
in the flavour sector at the precision frontier~\cite{O'Leary:2010af,Aushev:2010bq}. Included in this
program are:
\begin{itemize}
 \item tests of the CKM matrix at the 1\% level and searching for non-CKM sources
 of CP violation in  B decays
\item precision searches and measurements of 
$B\to K^{(*)} \ell^+ \ell^-$, $B\to\tau\nu$, and $B\to D^{(*)}\tau\nu$,
which all use the B-recoil technique
\item precision studies of the tau lepton, including searches for
 lepton-flavour-violating decays, measurements of g-2 and the electric dipole
moment of the tau, CP violation, the CKM element $|V_{us}|$
\item precision measurements of charm, including mixing and CP violation
\item searches for low mass Higgs bosons predicted in theories beyond the standard model
\item searches for evidence of a dark sector
\item if a polarized beam is available, precision electroweak physics that includes precise
and unique measurements of the neutral current vector coupling of b, charm, tau, muons and electrons
\end{itemize}
The program also includes precision QCD measurements involving  $\Upsilon(5S)$,
 ISR radiative return and hadron spectroscopy.

The physics motivation for the e$^+$e$^-$ super B factories is independent of
results from the  LHC: if LHC finds new physics, precision flavour input is
essential to further understand those discoveries. On the other hand,
 if the LHC finds no 
evidence for new physics, the  high statistics b, charm and tau samples
provide a unique way to probe for new physics beyond the TeV scale.

Regarding the interplay between e$^+$e$^-$ machines and LHCb: the two experiments
are highly complementary\cite{Bevan:2012hd}. LHCb will have high statistics samples of both B$_s$ and
B mesons  and will produce measurements that dominate the all-charged final states.
However, a super e$^+$e$^-$ machine will dominate B measurements of final states
 with $\pi^0$  or other neutral particles, including neutrinos. 
 The e$^+$e$^-$ program also includes extensive precision studies of the tau and a number
 of other non-flavour physics topics.

\subsection{Machines}
SuperB~\cite{Bona:2007qt,Biagini:2010cc} was designed to achieve a luminosity of 10$^{36}$cm$^{-1}$s$^{-1}$
by  having beams cross with a  large Piwinski angle (as at DAPHNE and KEKB);
 very low vertical and horizontal beta functions (as designed for the ILC); low horizontal and
 vertical emittances (as used in light sources); and ampere-level beam currents (as in PEP-II and KEKB).
For the $\Upsilon(4S)$ running, the positrons  were to have an energy of 6.7~GeV and the electrons,
polarized at the source, an energy of 4.2~GeV.

SuperKEKB~\cite{Abe:2010sj}
 is designed for  a luminosity of 8$\times$10$^{36}$cm$^{-1}$s$^{-1}$ with a large half crossing
angle of 41.5~mrad; horizontal emittances of 3.2~nm for the low energy (LER) and 5.0~nm high energy (HER)
 beam; beta functions at the IP of $\beta_x^*/\beta_y^*$ equal to 32~mm/0.27~mm (25~mm/0.31~mm) for the 
LER (HER); currents of 3.6~A (2.6~A) for the LER (HER); and a beam-beam tune shift of 0.0886 (0.0830) for the
 LER (HER).  The positron beam is 4~GeV and electron beam 7~GeV at SuperKEKB and the machine 
 does not have a polarized beam in its baseline design. The beams are injected into their respective
rings with low emittance: the electrons are produced using a low emittance gun, whereas
the positron beam uses a damping ring to ensure low emmittance.

\subsection{Detectors}
The SuperB detector~\cite{Grauges:2010fi} 
 at the Cabibbo Lab was to reuse many components from BaBar: the magnet,
DIRC bars and barrel CsI(Tl) calorimeter. The new components were to include:
a new silicon detector with an additional `Layer 0' around a smaller beam pipe; a new small cell drift chamber
with the possibility of incorporating cluster counting in order to improve particle 
identification; a new way to read out the DIRC that does not project through a water tank
as was done in BaBar;
 a new forward calorimeter; and a new muon detection system. SuperB was entertaining the
 possibility of a  forward particle identification system and had plans to incorporate
 a backward calorimeter into the detector.

The Belle~II detector~\cite{Abe:2010sj}  reuses the magnet and barrel and endcap calorimeters
 from Belle. It will have a new vertex detector with four layer double-sided silicon and two DEPFET pixel layers
close to the beam; a small cell, larger radius drift chamber; a barrel particle identification system
employing a novel technology,
called the Time of Propagation (TOP) system, which uses a quartz radiator and incorporates a combination of
Cherenkov radiation and propogation times to produce particle identification information~\cite{Nishimura:2010gj};
 and a new $K_L$-muon detector that will use scintillators read out with
 a multi-pixel photon counter (MPPC). In an upgrade, the CsI(Tl) endcap calorimeter crystals
 will be replaced with significantly faster pure CsI scintillators.
 The detector is scheduled to begin commissioning in 2015.

\subsection{Status}

Although, as mentioned, the Italian government cancelled the SuperB project at the end of 2012,
the status regarding SuperKEKB is excellent.
Approximately US\$100 million for the machine was 
approved in 2009 via Japan's Very Advanced Research Support Program.
 Full approval by the Japanese government came in December 2010. The project
was formally in the JFY2011 budget and approved by the Japanese Diet end of
March 2011. Most non-Japanese funding agencies have also already allocated
sizable funds for the upgrade of the detector.
Construction was started in 2010 and fortunately there was little damage
caused by the March 2011 earthquake - it did not introduce any delay in the project.
The ground breaking ceremony was held in November 2011 and both
 SuperKEKB and Belle~II construction is proceeding according to schedule.

\section{Summary}

BES~III has concrete plans for an exciting  physics program through 2015
and is scheduled to continue to run for another eight to ten years.

There are promising developments for a Super Charm-Tau Factory in Novosibirsk
with a sizable community associated with the project, although
the project is not yet approved for funding.

The  super B e$^+$e$^-$ flavour factories provide an extremely broad, rich, and
exciting physics program with sensitivity to new physics
that is complementary to the LHC. There is flexibility in ways that
such machines can achieve the high luminosity with beam currents and power comparable to
current facilities.
Unfortunately, the SuperB project was cancelled by the Italian government in late 2012.
On the other hand, SuperKEKB received Japanese Diet approval for the complete
project in 2011 and the construction on that project is proceeding well.

\bigskip
I am grateful to Xiaoyan Shen, Boris Shwartz, and Peter Krizan for providing
information about BES~III, Super Charm/Tau and Belle~II/SuperKEKB used in this
this review.

\end{document}